\documentclass[a4paper,onecolumn]{article}

\usepackage{graphicx}
\usepackage{amsmath, amssymb}

\newcommand{\me}{\mathrm{e}}
\newcommand{\mi}{\mathrm{i}}
\newcommand{\mca}{\mathcal{A}}
\newcommand{\mcl}{\mathcal{L}}
\newcommand{\uxu}{U(1) \otimes U(1)}

\DeclareMathOperator{\imag}{Im}

\begin{document}

\title{Stability of domain walls coupled to Abelian gauge fields}
\author{Damien P. George\thanks{d.george@physics.unimelb.edu.au}
\and Raymond R. Volkas\thanks{r.volkas@physics.unimelb.edu.au}}
\date{\normalsize \it School of Physics, Research Centre for High Energy
Physics,\\The University of Melbourne, Victoria 3010, Australia}

\maketitle

\begin{abstract}
Rozowsky, Volkas and Wali~\cite{rozowsky} recently found interesting
numerical solutions to the field equations for a gauged $\uxu$ scalar
field model.  Their solutions describe a reflection-symmetric domain
wall with scalar fields and coupled gauge configurations that
interpolate between constant magnetic fields on one side of the
wall and exponentially decaying ones on the other side.  This
corresponds physically to an infinite sheet of supercurrent
confined to the domain wall with a linearly rising gauge potential
on one side and Meissner suppression on the other.
While it was shown that these static solutions satisfied the field
equations, their stability was left unresolved.  In this paper, we
analyse the normal modes of perturbations of the static solutions
to demonstrate their perturbative stability.
\end{abstract}


\section{Introduction}

Topological and non-topological defects are interesting classes of
solutions to study for a large range of physical systems.  They
are frequently manifest in cosmological models described by
classical relativistic field equations and show up in quantised
systems as non-perturbative effects.  The more specific kind of
defect known as a domain wall or kink can act as an interface,
separating two regions described by the same physics but with
different boundary conditions.  Condensed matter physics uses
domain walls to model phase transitions and large scale structures
in a system.  Additionally, domain wall solutions are used as a
basis for brane-world
models~\cite{add1,rs1,rs2,akama,rubakov,antoniadis},
where our universe is embedded in a higher dimensional space.
This space is described throughout by the same physical model, but
different asymptotic vacuum behaviour in the higher dimension
induces a kink defect to which our $3+1$ dimensional world is
confined.

In one such brane-world toy model, Rozowsky, Volkas and
Wali~\cite{rozowsky} have found interesting solutions consisting
of a pair of concentric domain walls coupled to a pair of $U(1)$
gauge fields.  Their solutions look physically like infinite
sheets of supercurrent confined to the wall, producing a linearly
increasing gauge potential on one side and Meissner suppression on
the other.  The purpose of this paper is to demonstrate the
perturbative stability of this configuration.  Our method is based
on an analysis of the normal modes of perturbations of the static
solutions, and we find that for a large range of parameters these
modes are oscillatory, remaining bounded in time.

In Section~\ref{sec:model} we present the model with a slight
generalisation and display the static solutions for light- and
space-like gauge fields.  Our stability analysis method is
outlined in Section~\ref{sec:stab} followed by a full
investigation demonstrating the perturbative stability of the
static configurations.  We give a summary of our results in
Section~\ref{sec:conc}.

\section{The model}
\label{sec:model}

We present a slight generalisation of the Rozowsky et al.\ model
which specifies two scalar fields $\phi_i$ ($i=1,2$) each with
independent local $U(1)$ gauge symmetries with associated gauge
fields $A_i^\mu$.  There is an additional discrete $\mathbb{Z}_2$
symmetry interchanging $\phi_1 \leftrightarrow \phi_2$ and
$A_1^\mu \leftrightarrow A_2^\mu$.  This model is a toy model
invented to study the clash-of-symmetries mechanism in its
simplest non-trivial setting; see~\cite{toner,shin,demaria,dando}.
A quartic scalar potential couples the scalar fields to each other
and permits domain wall solutions asymptoting to different
degenerate minima.  The Lagrangian density is
\begin{equation}
\label{eq:lag}
\mcl = (D^\mu \phi_1)^* D_\mu \phi_1 + (D^\mu \phi_2)^* D_\mu \phi_2
    - \tfrac{1}{4} F_1^{\mu \nu} F_{1 \mu \nu} - \tfrac{1}{4} F_2^{\mu \nu} F_{2 \mu \nu}
    - V(\phi_1, \phi_2),
\end{equation}
with the scalar field potential given by
\begin{equation}
\label{eq:pot}
V(\phi_1, \phi_2) = \lambda_1 ( \phi_1^* \phi_1 + \phi_2^* \phi_2 - v^2)^2 + \lambda_2 \phi_1^* \phi_1 \phi_2^* \phi_2.
\end{equation}
We work in the $\lambda_{1,2} > 0$ parameter regime, where the
degenerate global minima are manifestly given by
\begin{equation*}
|\phi_1| = v, \quad \phi_2 = 0 \quad \mbox{and} \quad \phi_1 = 0, \quad |\phi_2| = v.
\end{equation*}

The $U(1)$ gauge fields $A_i^\mu$ are described in the usual way
by $F_i^{\mu \nu}$ and their appearance in the covariant
derivative
\begin{equation*}
D^\mu = \partial^\mu - \mi Q_1 A_1^\mu - \mi Q_2 A_2^\mu,
\end{equation*}
where $Q_i$ is the charge operator associated with the $U(1)_i$
symmetry.  Keeping the discrete $\mathbb{Z}_2$ symmetry, the
$U(1)_1 \otimes U(1)_2$ charges of the scalar fields are
$\phi_1 \sim (e, \tilde{e})$ and $\phi_2 \sim (\tilde{e}, e)$,
with $e$ and $\tilde{e}$ constants.\footnote{The Rozowsky et
al.\ model took $\tilde{e}=0$, so this is a slight
generalisation, first introduced in~\cite{dando}.}

The Euler-Lagrange equations of motion for the scalar and gauge
fields are
\begin{equation}
\label{eq:uxu-general}
\begin{aligned}
D^\mu D_\mu \phi_i &= -2 \lambda_1 \phi_i (\phi_i^* \phi_i + \phi_j^* \phi_j - v^2) - \lambda_2 \phi_i \phi_j^* \phi_j, \\
\partial_\nu F_i^{\nu \mu} &= 2 \imag \left( e \phi_i^* D^\mu \phi_i + \tilde{e} \phi_j^* D^\mu \phi_j \right),
\end{aligned}
\end{equation}
where $i = 1$ and $j = 2$, or $i = 2$ and $j = 1$ (this notation
is to be understood in subsequent equations).
Following \cite{rozowsky}, we look for static solutions that
depend only on $z$ and utilise a polar decomposition for the
scalar fields; $\phi_i(z) = R_i(z) \me^{\mi \Theta_i(z)}$.
The scalar potential~\eqref{eq:pot} allows one to construct domain
wall solutions by requiring $\phi_i$ to asymptote to
different degenerate minima of $V$ in an appropriate limit.  Since
our fields vary only with $z$, this will be the dimension
perpendicular to the wall and the limit will be as $|z|$ tends to
infinity.  This gives us the boundary conditions
\begin{equation}
\label{eq:bc}
\begin{aligned}
|\phi_1| \rightarrow 0, \quad |\phi_2| \rightarrow v \quad &\mbox{as} \quad z \rightarrow -\infty, \\
|\phi_1| \rightarrow v, \quad |\phi_2| \rightarrow 0 \quad &\mbox{as} \quad z \rightarrow \infty,
\end{aligned}
\end{equation}
or vice-versa.

Since we are working with gauge fields we have the freedom to
choose two gauges, one for each $A_i^\mu$; the Lorentz gauge
$\partial_\mu A_i^\mu = 0$ turns out to be the most suitable
choice for both.  The algebra simplifies if instead of $A_i^\mu$
one considers the linear combination
$\mca_i^\mu = e A_i^\mu + \tilde{e} A_j^\mu$.
Working with these choices, the field equations of
motion~\eqref{eq:uxu-general} reduce to
\begin{align}
\label{eq:stat-scalar}
R_i''                &= -R_i ( {\mca_i^t}^2 - {\mca_i^x}^2 - {\mca_i^y}^2 )
                        + 2 \lambda_1 R_i (R_i^2 + R_j^2 - v^2) + \lambda_2 R_i R_j^2, \\
\label{eq:stat-gauge}
{\mca_i^{(t,x,y)}}'' &= 2 (e^2 + \tilde{e}^2) R_i^2 \mca_i^{(t,x,y)} + 4 e \tilde{e} R_j^2 \mca_j^{(t,x,y)}, \\
\label{eq:stat-gauge-z}
{\mca_i^z}'          &= 0, \\
\label{eq:stat-phase}
\Theta_i'            &= -\mca_i^z,
\end{align}
where prime denotes differentiation with respect to $z$.  We
immediately see that the $\mca_i^z$ and hence the $A_i^z$ are pure
gauge and do not contribute to the physics; neither do the
$\Theta_i$.

To further simplify the problem, we note that each gauge component
$\mca_i^{(t,x,y)}$ exhibits the same dynamics
in~\eqref{eq:stat-gauge} and appears quadratically
in~\eqref{eq:stat-scalar}.  Thus, the qualitative physical
behaviour depends on whether the gauge field configuration is
space-like, time-like or light-like.  Expressing this behaviour as
the single field $\mca_i$ we have
\begin{align}
\label{eq:stat-r}
R_i''    &= k R_i \mca_i^2 + 2 \lambda_1 R_i (R_i^2 + R_j^2 - v^2) + \lambda_2 R_i R_j^2, \\
\label{eq:stat-a}
\mca_i'' &= 2 (e^2 + \tilde{e}^2) R_i^2 \mca_i + 4 e \tilde{e} R_j^2 \mca_j,
\end{align}
where $k = +1, \, 0, \, -1$ for space, light- and time-like
gauge fields respectively.  One can see that
equation~\eqref{eq:stat-r} is consistent with the boundary
conditions~\eqref{eq:bc} so long as $k \ge 0$.  For
$k = -1$ the asymptotic behaviour of $R_i$ is oscillatory and
so we discard this time-like scenario.

Considering equation~\eqref{eq:stat-a} on the side of the wall
where $R_i \rightarrow v$ and $R_j \rightarrow 0$, we see that
$\mca_i'' \rightarrow 2 (e^2 + \tilde{e}^2) v^2 \mca_i$.  Since
physical solutions must be bounded, we conclude that $\mca_i$ is
exponentially suppressed.  On the other side of the wall,
$R_i \rightarrow 0$, $R_j \rightarrow v$ and, using the previous
result, $\mca_j \rightarrow 0$.  Thus $\mca_i'' \rightarrow 0$ and
this gauge field takes on a linear form.  We note that all
measurable quantities associated with $\mca_i$ arise through
derivatives and so this unbounded solution is still physical.

The set of equations~\eqref{eq:stat-r} and~\eqref{eq:stat-a}
cannot, in general, be solved analytically and we use the
relaxation-on-a-mesh technique to obtain numerical solutions.

\begin{figure}
\centering
\includegraphics[width=0.49\textwidth]{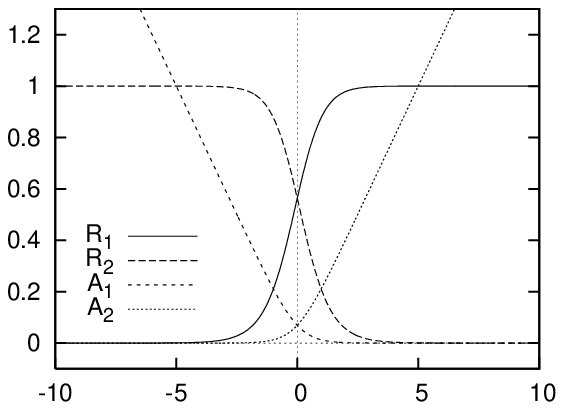} \hfill
\includegraphics[width=0.49\textwidth]{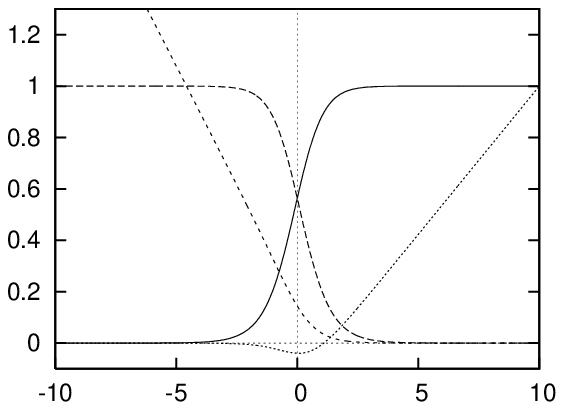}
\includegraphics[width=0.49\textwidth]{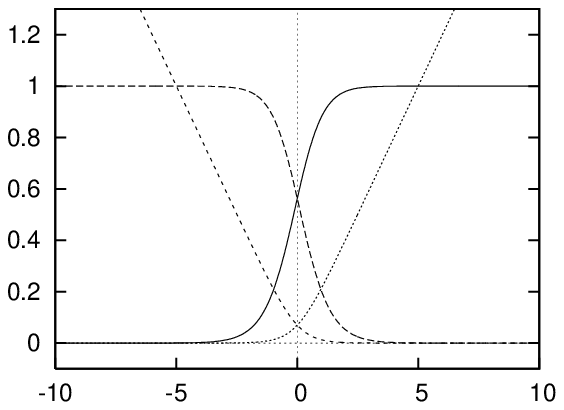} \hfill
\includegraphics[width=0.49\textwidth]{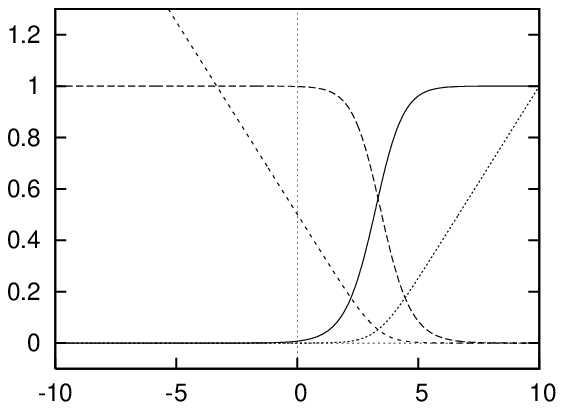}
\caption{
\small
Static solutions for the two scalar and two gauge fields in the
$\uxu$ model, plotted against $z$.  The top two plots are for the
light-like case, the bottom two for the space-like case.  All
plots have $e=1$, $\tilde{e}=\tfrac{1}{2}$, $\lambda_1=1$,
$\lambda_2=2$ and $v=1$.  The plots on the left have symmetric
boundary conditions for the gauge fields, those on the right
asymmetric.  In the light-like case, the scalar fields do not feel
the gauge fields and thus do not depend on the choice of gauge field
boundary conditions.  This is unlike the space-like case where the
scalar fields centre on the gauge fields to restore the reflection
symmetry.
}
\label{fig:stat}
\end{figure}

In the light-like case, $k = 0$, and $R_i$ can be solved for
independently of $\mca_i$.  Typical solutions are shown in the top
two plots in Figure~\ref{fig:stat}.  The scalar fields assume a
typical domain wall configuration asymptoting to distinct minima
of the potential.  As the boundary conditions for the $R_i$ are
symmetric under $z$ reflection, the solutions for these scalar
fields are just reflections of each other.  In the left plot, the
boundary conditions for the two gauge fields are also reflection
symmetric.  In the right plot, a different boundary condition is
used for $\mca_2$.  As the scalar fields do not feel the presence
of the gauge fields they have exactly the same solutions in both
cases.

For the space-like case, $k = 1$, and the scalar and gauge
fields are fully coupled.  Solutions are shown in the bottom two
plots in Figure~\ref{fig:stat} with all parameters, except
$k$, mimicking the top two plots.  Although they look similar,
the light- and space-like plots on the left are slightly
different.  A more significant difference between these two
scenarios is evident in the right plots where the boundary
conditions for the two gauge fields are different.  In the
space-like case the scalar fields are influenced by the gauge
fields and the favourable configuration is that with exact
reflection symmetry.  The boundary conditions serve to simply
shift the centre of the domain wall and the right plot on the
bottom is an exact translation of the left plot.  Our result is
contrary to the claim in~\cite{rozowsky} that asymmetric
boundary conditions in the space-like case are not equivalent to
spatial translations of the domain wall centre.

Disregarding the technical details, the qualitative features of
this $\uxu$ model are the reflection symmetric scalar fields in a
domain wall configuration and the partially suppressed gauge
fields.  This suppression of $\mca_i$ under their respective $R_i$
is physically similar to the Meissner effect and serves to
semi-localise the gauge fields.  We make the physical
interpretation of an infinite sheet of supercurrent confined to
the wall, producing a constant magnetic field in the region
opposite the suppression.

While we have shown the existence of static solutions to the
model described by~\eqref{eq:lag}, we have not established their
stability.  In the next section we demonstrate that under small 
perturbations, the solutions to equations~\eqref{eq:stat-r}
and~\eqref{eq:stat-a} are stable.

\section{Stability}
\label{sec:stab}

The essence of static solutions is their time independence, but a
physical processes requires the underlying fields to evolve in
time.  We must thus ensure that the static configurations found in
the previous section are not destroyed by time-dependent
perturbations.  In this section we add to the static solutions
perturbations expressed as normal modes and arrive at a set of
equations characterising these normal eigenfunctions and
associated eigenvalues.  We then demonstrate that the eigenvalues
are all positive and hence the perturbations are oscillatory.

We begin by taking each static field, including all four gauge
components $\mca_i^\mu$, and adding a perturbation factored as an
unspecified spatial part and a time dependent complex
exponential.  This exponential represents an arbitrary normal mode
of the perturbation, characterised by an eigenvalue which is in
general complex.  We express this construction in the
substitutions
\begin{equation}
\label{eq:pert-sub}
\begin{aligned}
R_i        &\rightarrow R_i(z) + r_i(z) \me^{\mi \omega_r t}, \\
\mca_i^\mu &\rightarrow \mca_i^\mu(z) + a_i^\mu(z) \me^{\mi \omega_a t}, \\
\Theta_i   &\rightarrow \Theta_i(z) + \theta_i(z) \me^{\mi \omega_\theta t}.
\end{aligned}
\end{equation}

By the choice of an explicitly complex exponential, if $\omega$ is
purely real then the perturbation will be oscillatory and hence
remain bounded in time.  On the other hand, if $\omega$ has an
imaginary component, the exponential will blow up, signifying
instability of the original static solution.

We take the original field equations~\eqref{eq:uxu-general}, make
the substitutions given by~\eqref{eq:pert-sub} and simplify using
the equations~\eqref{eq:stat-scalar}, \eqref{eq:stat-gauge},
\eqref{eq:stat-gauge-z} and ~\eqref{eq:stat-phase} for the static
fields.  We work to first order in $r_i$, $a_i^\mu$ and $\theta_i$
and consider only independent perturbations, which decouples the
resulting set of equations to give
\begin{align}
\label{eq:pert-scalar}
\left( -\partial_z^2 - ( {\mca_i^t}^2 - {\mca_i^x}^2 - {\mca_i^y}^2 ) + 2 \lambda_1 (3 R_i^2 + R_j^2 - v^2) + \lambda_2 R_j^2 \right) r_i &= \omega_r^2 r_i, \\
\label{eq:pert-gauge}
\left( -\partial_z^2 + 2 (e^2 + \tilde{e}^2) R_i^2 \right) a_i^\mu &= \omega_a^2 a_i^\mu, \\
\label{eq:pert-phase}
\left( -\partial_z^2 - 2 \frac{R_i'}{R_i} \partial_z \right) \theta_i &= \omega_\theta^2 \theta_i.
\end{align}

Before we continue with these equations, we must first establish
a general result.  Given the equation
\begin{equation*}
f''(z) + V(z) f'(z) + W(z) f(z) = 0,
\end{equation*}
one can show that if $W(z) < 0$ for all $z$, then there exist no
non-trivial solutions for $f(z)$ on the domain
$z \in \mathbb{R}$ with $f(z) \rightarrow 0$ as
$|z| \rightarrow \infty$.  To see this consider large $-z$ with
$f$ taking a vanishingly small positive value.  For non-trivial
solutions, $f$ must increase as $z$ increases\footnote{Since
$f(-\infty) = 0$ there must be a region where $f$ increases if
it is to attain a finite positive value.} and so $f' > 0$.
For solutions where $f$ becomes vanishingly small for large $z$,
we require $f' < 0$ for some subsequent region of the z-axis.
This change in the sign of the first derivative requires $f'' < 0$
for some region, in particular we must have $f'' < 0$ when
$f' = 0$, i.e.\ at the turning point.  But at this point we have
$f'' = -W(z) f$ and since $f > 0$ and $W(z) < 0$ for all $z$ we
have $f'' > 0$.  Thus the function is positive with a positive
gradient and can never turn back towards the z-axis.  A similar
argument holds when $f$ is below the axis; it can never turn back
up.  Hence there are no non-trivial bounded solutions if
$W(z) < 0$ for all $z$.

We now return to the issue of stability.  Consider
equation~\eqref{eq:pert-phase} with $f(z) = \theta_i(z)$ and
$W(z) = \omega_\theta^2$.  If $\omega_\theta^2 < 0$ then one would
have $W(z) < 0$ for all $z$ and by the previous result the only
solution for $\theta_i$ would be the trivial one.  Thus there are
no negative eigenvalues for equation~\eqref{eq:pert-phase} with
bounded eigenfunctions $\theta_i$.  Note that the condition that
eigenfunctions $\theta_i$ be bounded does not preclude the
analysis of the bounded nature of the perturbations.  The
perturbation to the static field is given in full by
$\theta_i(z) \me^{\mi \omega_\theta t}$ where by definition of a
perturbation, $\theta_i(z)$ must be small and bounded.  It is the
nature of the temporal part $\me^{\mi \omega_\theta t}$, hence the
eigenvalues, that determines if the fields are stable.  Since we
have shown that $\omega_\theta^2 \ge 0$ we have $\omega_\theta$
real and thus oscillatory perturbations and hence a stable static
field $\Theta_i$.

For the gauge fields, inspection of equation~\eqref{eq:pert-gauge}
yields
\begin{equation*}
W(z) = \omega_a^2 - 2 (e^2 + \tilde{e}^2) R_i^2.
\end{equation*}
For bounded $a_i^\mu(z)$ we require $W(z) \ge 0$ for some non-zero
domain of $z$.  This means that we need
\begin{equation*}
\omega_a^2 \ge 2 (e^2 + \tilde{e}^2) R_i^2
\end{equation*}
for some $z$.  Since $\omega_a^2$ is a constant it must be greater
than or equal to the minimum of $2 (e^2 + \tilde{e}^2) R_i^2$,
hence non-negative.  Thus we have shown that the static gauge
fields are stable under small time dependent perturbations.

\begin{figure}
\centering
\includegraphics[width=0.49\textwidth]{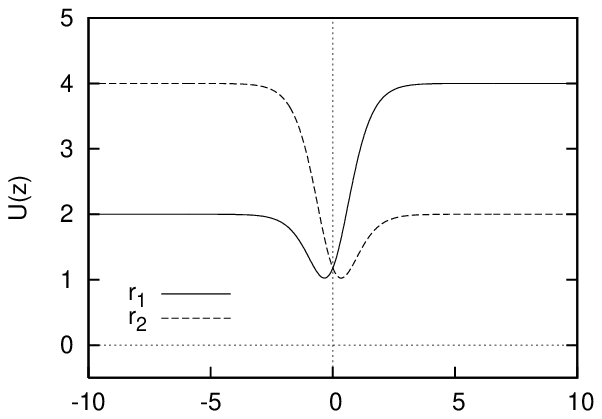} \hfill
\includegraphics[width=0.49\textwidth]{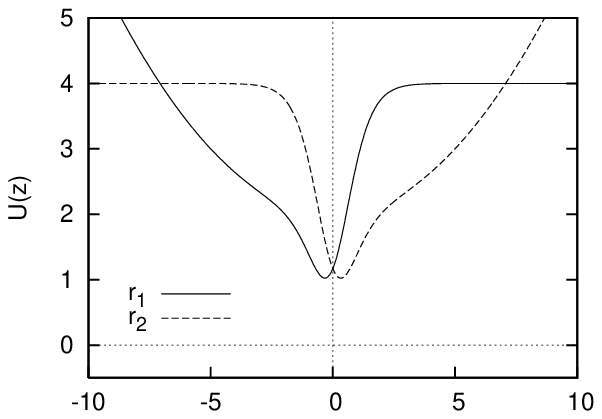}
\caption{
\small
The function $U(z)$ used to determine the eigenvalues of the $r_i$
perturbation in the light- (left) and space- (right) like cases,
plotted as a function of $z$.  The parameters and corresponding
field configurations are as in the reflection symmetric cases in
Figure~\ref{fig:stat}.  There are two plots in each graph
corresponding to $U(z)$ with $i=1$, $j=2$ and $i=2$, $j=1$.
}
\label{fig:pert-pot}
\end{figure}

Following a similar argument for the scalar fields,
equation~\eqref{eq:pert-scalar} gives us the bound on the
eigenvalues as
\begin{equation*}
\omega_r^2 \ge \min(U(z)),
\end{equation*}
where
\begin{equation}
\label{eq:u}
U(z) = -( {\mca_i^t}^2 - {\mca_i^x}^2 - {\mca_i^y}^2 ) + 2 \lambda_1 (3 R_i^2 + R_j^2 - v^2) + \lambda_2 R_j^2.
\end{equation}
It is not so clear as to the sign of this function.  We analyse
the light-like case first where the $\mca_i^\mu$ terms are absent.
In this case, as can be seen from equation~\eqref{eq:stat-r} with
$k = 0$, the scalar field configuration and hence $U(z)$
depend only on the parameters $\lambda_1$, $\lambda_2$ and $v$.
Since $v$ can be absorbed into a rescaling of the $R_i$, we only
have two parameters to consider.  A typical plot of the function
$U(z)$ for the two permutations of $i$ and $j$ is shown on the
left in Figure~\ref{fig:pert-pot}.

\begin{figure}
\centering
\includegraphics[width=0.49\textwidth]{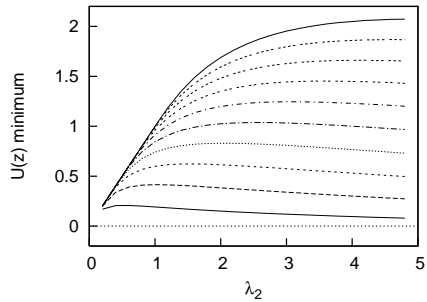}
\caption{
\small
The minimum of the function $U(z)$ in the light-like case plotted
against $\lambda_2$.  The upper curves correspond to successively
larger values of $\lambda_1$, which runs from $0.2$ to $2$ in
steps of $0.2$.
}
\label{fig:li-pert-min}
\end{figure}

We see that $U(z) > 0$ for this choice of parameters.
Figure~\ref{fig:li-pert-min} shows the minimum of $U(z)$ for a
large range of values of $\lambda_1$ and $\lambda_2$.  Since all
minima are positive, it must be that $\omega_r^2 > 0$ and hence
the static scalar fields in the light-like case are stable, at
least for this range of parameters.

\begin{figure}
\centering
\includegraphics[width=0.49\textwidth]{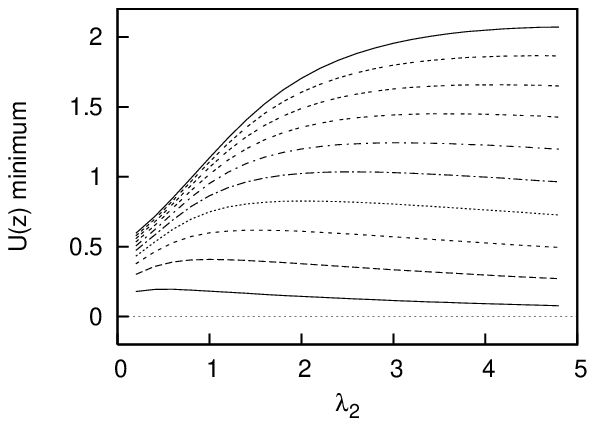} \hfill
\includegraphics[width=0.49\textwidth]{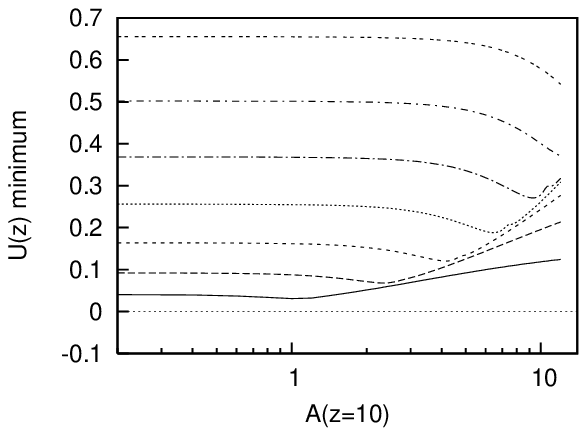}
\caption{
\small
The minimum of the function $U(z)$ in the space-like case.
The plot on the left is against $\lambda_2$ with upper curves
corresponding to larger values of $\lambda_1$, which runs from
$0.2$ to $2$ in steps of $0.2$.  The plot on the right is
against the boundary condition for the gauge field with upper
curves corresponding to larger values of $v$, which runs from
$0.2$ to $0.8$ in steps of $0.1$.
}
\label{fig:sp-pert-min}
\end{figure}

In the space-like scenario, the results are similar.  The plot in
the right of Figure~\ref{fig:pert-pot} shows equation~\eqref{eq:u}
with $\mca_i^x$ and $\mca_i^y$ present.
Figure~\ref{fig:sp-pert-min} shows the minimum of $U(z)$ for
various values of the parameters and boundary conditions for the
gauge fields.  It is clear that the minima are all positive and
so the static scalar field configuration is also stable in the
space-like case.

Since each field in the model permits static solutions which are
independently stable, we conclude that the static configuration as
a whole is a stable one.  We have also verified this analysis with
explicit numerical calculation of the eigenvalues.

\section{Conclusion}
\label{sec:conc}

Static solutions to a $\uxu$ gauged scalar model were recently
found by Rozowsky, Volkas and Wali~\cite{rozowsky}.  In this paper
we have generalised the model slightly, presented the static
solutions and demonstrated the stability of this field configuration.
We achieved this by adding small time dependent perturbations, in the
form of normal modes, to the static fields and obtaining
eigenvalue equations.  It was shown that these eigenvalues,
corresponding to the normal modes, were positive for a large range
of parameters in the model.  Thus the perturbations were
oscillatory and the static fields stable.

\section*{Acknowledgments}

DPG was supported by the Puzey bequest to the University of
Melbourne and RRV by the Australian Research Council.

\bibliography{stab-wall-gauge}

\end{document}